\documentclass{PoS}

\title{A Survey of Active Galaxies with the HAWC Gamma-ray Observatory}

\ShortTitle{HAWC AGN survey}

\author{\speaker{Alberto Carrami\~nana}, Daniel Rosa González, Sara Couti\~no de Le\'on, Anna Lia Longinotti for the HAWC Collaboration\thanks{AC acknowledges the support of CONCyTEP (Consejo de Ciencia y Tecnolog\'{\i}a del Estado de Puebla) that made possible his participation at the ICRC~2019. 
For the collaboration list see PoS(ICRC2019)1177. 
Further acknowledgements and the complete list of authors are available at  https://www.hawc-observatory.org/collaboration/icrc2019.php}\\
        Instituto Nacional de Astrof\'{\i}sica, \'Optica y Electr\'onica\\ Luis Enrique Erro 1, Tonantzintla, Puebla, M\'exico \\
        E-mail: \email{alberto@inaoep.mx}}

\abstract{The High Altitude Water Cherenkov  Gamma-Ray Observatory has been accumulating a progressively deeper exposure of the TeV sky since its inauguration in March 2015. Located at geographical latitude $+19^\circ$N, HAWC has been able to perform a deep and unbiased survey of two thirds of the sky. We analyzed three years of HAWC data searching for long term persistent emission from a redshift limited ($z\leq 0.3$) sample of active galactic nuclei drawn from the {\it Fermi}-LAT 3FHL catalog. The HAWC dataset confirms the high significance detection of the two nearest BL Lac objects, \mbox{Mrk 421} and \mbox{Mrk 501}, and sets limits for the rest of the sample, down to integrated photon fluxes of order $N(>0.5~{\rm TeV})\lesssim 10^{-12}\,\rm cm^{-2}s^{-1}$. We present and discuss some of the results of this survey, focusing on individual objects of particular interest.}

\FullConference{36th International Cosmic Ray Conference -ICRC2019-\\
		July 24th - August 1st, 2019\\
		Madison, WI, U.S.A.}

\begin{document}

\section{Introduction: active galatic nuclei}
Active galactic nuclei (AGN) are powerful extragalactic persistent sources of radiation throughout the electromagnetic spectrum. Although the majority of AGN are radio-quiet, the intense radio emission of quasars facilitated their discovery and the generic interpretation of AGN in terms of a massive central engine~\cite{agn-book,hf63}. 
The vast amounts of energy radiated in relatively small regions led to the current standard supermassive black hole scenario, in which the central compact object with a mass up to $\gtrsim 10^{9}\,\rm M_\odot$ accretes matter with a rate that may be in excess of $1\rm M_\odot/\rm year$. The Schwarzschild scale was introduced to account for the very rapid variability observed in particular classes of AGN~\cite{salpeter64,zeldovich64}. The presence of angular momentum forces accretion to proceed via dissipative disks, establishing an axial symmetry consistent with the commonly observed jets and outflows~\cite{alpha-disk,agn-jets}. 
The advent of space telescopes in orbit led to the discovery of MeV-GeV emission in radio-loud AGN and the formal identification of $\gamma$-ray sources with compact flat-spectrum radio sources~\cite{3c273-cosb,agn-egret}. It is established that $\gamma$-ray emitting AGN are predominantly blazars for which the jet axis matches our line of sight~\cite{blazars,agn-araa}. 

In addition to their inherent multi-wavelength character, there is mounting evidence of AGN as multi-messenger sources. They have been tagged as some of the best candidate sources for cosmic rays, up to the ultra-high-energy range~\cite{agn-cr}; the recent positional match of the neutrino event IceCube-170922A with the $\gamma$-ray emitting BL Lac object TXS~0506+056 provided strong evidence in that respect~\cite{agn-neutrino}. As AGN have been studied in the very high energy (VHE; $\geq 100\,\rm GeV$) $\gamma$-ray regime mostly through short pointed observations with atmospheric Cherenkov telescopes (ACTs), we lack a clear picture of their baseline behavior, characterized by long term continuous observations. Two limitations for conducting all-sky AGN surveys in the VHE range have been the lack of air shower arrays with the sensitivity required to detect extragalactic sources; and 
the redshift limits imposed by photon-photon absorption with the extragalactic background light (EBL), which renders the Universe opaque to TeV gamma rays from distant sources. 

The High Altitude Water Cherenkov (HAWC) observatory has been monitoring two-thirds of the sky almost continuously since its formal inauguration on the Spring equinox day of 2015. With its instrumental response peaking at a few TeV, the HAWC observatory has the sensitivity to perform a follow-up survey of AGN detected by {\em Fermi}-LAT within the redshift range $z\leq 0.3$. The first results of this study are presented in this paper.


\section{The HAWC Gamma-Ray Observatory\label{hawc-obs}}
HAWC is a panoramic TeV gamma-ray detector located in Sierra Negra, in the Mexican State of Puebla, at an altitude of 4100~m and geographical latitude of 19$^\circ$N. It consists of a dense array of 300 individual water Cherenkov detectors (WCDs), each 5m tall, 7.2m in diameter and instrumented with four photomultiplier tubes - three of 8-inches, one of 10-inches. Each WCD holds 180~m$^3$ of purified water isolated from the ambient light. The array covers a physical area of about 22,500~m$^2$ and has operated with a duty cycle $\gtrsim 95\%$ since starting full-operations mode in March 2015. With a field-of-view of 1.8~sr, HAWC scans every sidereal day two-thirds of the sky with the sensitivity to detect the Crab Nebula at 5$\sigma$ per transit~\cite{hawc-crab}.

The HAWC observatory detects cosmic particles with a $\gtrsim 20\,\rm kHz$ rate. Data are registered event per event, keeping timing and charge deposit information for each of the 1200 channels. The current standard analysis separates the data according to $f_{hit}$, the fraction of channels recording a signal, using nine $f_{hit}$ bins. The median energy of the events increases with increasing $f_{hit}$ bin number $\cal B$, while the point-spread-function becomes narrower with increasing ${\cal B}$. In particular we note that the distribution of events in the first bin peaks at 0.5~TeV~\cite{hawc-crab}. While the correspondence is rough, such that a 5~TeV photon has sizable ($\geq 5\%$) probabilities of being registered in any bin between ${\cal B}=1$ and 6, the overall trend allows to use $\cal B$ as an energy proxy.  
Gamma-hadron separation is made on basis of the morphology of the charge deposits, with specific cut parameters defined for each bin. At least half the photons pass the standard cuts in any bin, while the fraction of hadrons rejected goes from $\sim 85\%$ in ${\cal B} =1$, to $\sim 99.9\%$ for ${\cal B}\geq 6$. Further details on the performance of HAWC can be found in the Crab Nebula validation observations and in the 2HWC catalog papers~\cite{hawc-crab,2hwc}.

\section{Photon-photon absorption by the EBL}
\begin{figure}
\includegraphics[width=\textwidth]{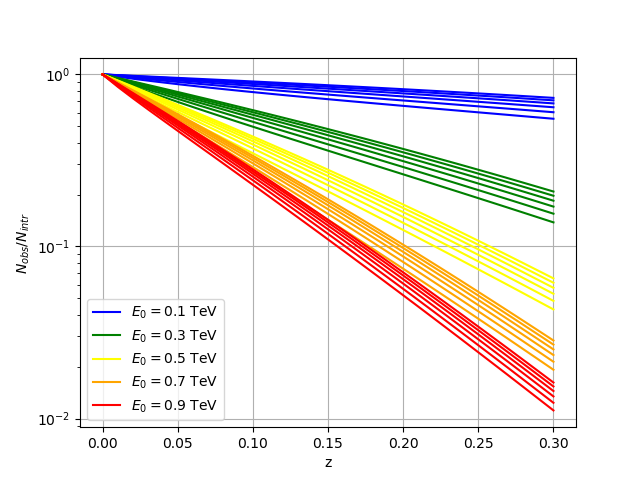}
\caption{The exponential dependence on redshift of $N_{obs}/N_{intr}$ for different values of the starting energy $E_0$ (in color) and the spectral index $\alpha$, which ranges from 2.0 (lower for each $E_0$) to 3.0 (upper for each $E_0$). The yellow lines correspond to $E_0=0.5\,\rm TeV$ which best describes the HAWC data.} \label{f1}
\end{figure}

The main limiting factor for the survey of AGN presented here is the absorption of TeV gamma rays due to the $\gamma\gamma\to {\rm e}^{+}{\rm e}^{-}$ process with photons from the EBL. This process is most efficient for absorbing $\gamma$ rays of energy $E$ just above its threshold, \mbox{$E h\nu \gtrsim \left(m_e c^{2}\right)^2\simeq 0.25\,\rm TeV\cdot eV ,$} given an intervening EBL photon of frequency  $\nu$. The opacity of the EBL to gamma rays depends on the photon energy $E$ and the distance to its source, often measured by the redshift $z$. This optical depth $\tau(E,z)$ is computable when the dependence of the EBL photon density with redshift, $n_\nu(z)$, is known. Different EBL models have been used to compare the predicted $\tau$ with observations. In this work we used the EBL model of~\cite{ebl-model} to compute the attenuation of simple power-law spectra integrated from an energy $E_0$.
We found that the observed photon flux, $N_{obs}$, follows an exponential dependence on redshift $z$,
\begin{equation}
N_{obs}(>E_0)=e^{-z/z_h} N(>E_0) , \label{horizon}
\end{equation}
where $N(>E_0)\propto E_{0}^{-\alpha+1}/(\alpha -1)$ is the integral of a power law with spectral index $\alpha$. The redshift scale $z_h$ is strongly dependent on $E_0$ and weakly dependent on $\alpha$, as illustrated in the Figure~\ref{f1} that asserts the exponential dependence on redshift. 
The exponential character of this function is inherited from the $e^{-\tau(E,z)}$ EBL attenuation.
HAWC data are consistent with $E_0=0.5\,\rm TeV$, which gives $z_h=0.108$ for $\alpha=2.5$, and a range from $z_h=0.095$ at $\alpha=2.0$ to $z_h=0.118$ at \mbox{$\alpha=3.0$}. 
The upper bound on the redshift for this follow-up survey of high energy {\em Fermi}-LAT sources, $z\leq 0.3$,  
was set consistently with the $E_0$ estimate.

\section{The 3FHL follow-up sample \label{muestra}}
We constructed our follow-up sample from the Third Catalog of Hard {\em Fermi}-LAT sources (3FHL), that contains 1556 objects detected in the energy range 10~GeV--2~TeV, with LAT data taken between August 2008 and August 2015~\cite{3fhl}. We selected AGN from the 3FHL catalog accesible from the HAWC site, i.e. culminating within $40^\circ$ of the local zenith, and that have a known redshift satisfying $z\leq 0.3$. These conditions narrow the list from 1231 AGN in the 3FHL to the 138 selected sources constituting our follow-up sample. The sample is composed of: 1 starburst galaxy (NGC~1068), 6 radiogalaxies, 117 BL Lac objects, 6 flat-spectrum-radio-quasars and 8 blazars of uncertain nature (``bcu'' following 3FHL terminology). 
In order to flag the best candidates we implemented three procedures:
\begin{enumerate}
\item We identified AGN with $TS \geq 10$ in all five 3FHL energy intervals, in particular \mbox{(0.5--2.0~TeV),} where $TS=2\Delta \log{\cal L}$ is the Test Statistic defined from the likelihood ($\cal L$) ratio between a source hypothesis and a null hypothesis in~\cite{3fhl}. 
Only two AGN, Mrk~421 and Mrk~501, have $\sqrt{TS}\geq 5$ in all LAT bands, with five other sources (I~Zw~187=1ES~1727+502, 1H~1013+498, B3~2247+381, RX~J0648.7+1516, 1H~1914--194) showing weak detections in the top LAT energy band.
\item We computed expected photon fluxes, $N_{obs}(>0.5\,{\rm TeV})$, extrapolating the 3FHL best fit spectral models including $\gamma\gamma$ attenuation by the EBL in the extrapolation. Most of the 3FHL spectral models are power-laws that allow to approximate EBL attenuation through expression~(\ref{horizon}). 
Four of the seven sources mentioned in the paragraph above were above 30~mCrab, set as a rough indicative threshold from the 5$\sigma$ sensitivity of HAWC of 1~Crab per transit and a time-span of $\sim 1000\,\rm days$. Under this criterion, M~87, IC~310, 1ES~2344+514 and TXS~0210+515 were added to the list of best candidates, increasing it to eleven objects.
\item The 3FHL catalog flags sources as detected or non-detected in the VHE range by ACTs (TeV=P, N flags respectively), defining a third candidate category (TeV=C) for non-detected sources with suitable GeV properties to be VHE emitters. Our sample contains 32 objects flagged as detected in the VHE range by ACTs. We extrapolated the ACT spectra of these to observed photon fluxes $N_{obs}(E>0.5\rm TeV)$. These extrapolations are taken as indicative, given that ACT observations can be biased to particular AGN activity states. The needed input was first gathered from the TeVCat~\cite{tevcat} and consulted more carefully in the publications referred by TeVCat. Keeping the 30~mCrab limit for the observed photon flux at $E\geq 0.5\,\rm TeV$, only one object was added: the intermediate redshift BL Lac H~1426+428, observed by HEGRA and VERITAS in the TeV range more than a decade ago~\cite{H1426+428-hegra,H1426+428-veritas}. 
\end{enumerate}

\section{Analysis and results}
\subsection{Analysis}
We present here 1017~days of effective HAWC data acquired between November 2014 and December 2017. The data are separated in the nine $f_{hit}$ bins described in section \S\ref{hawc-obs} for their analysis. We use standard maximum-likelihood analysis fitting a power-law spectrum assumed as intrinsic to each source, with the corresponding EBL attenuation at the redshift, and location, of the counterpart,
\begin{equation}
\left(dN / dE\right)_{int} = \phi_1 \left(E / 1~{\rm TeV}\right)^{-\alpha} \quad \Leftrightarrow\quad (dN/dE)_{obs} = \phi_1 \left(E/1~{\rm TeV}\right)^{-\alpha}\, e^{-\tau(E,z)} ;
\end{equation}
where $\phi_1$ is the differential photon flux normalization at 1~TeV and $\alpha$ the spectral index. The likelihood ratio between the best fit for a point source (${\cal L}_1$) and the null hypothesis  (${\cal L}_0$) is then computed. On a first run we fitted both $\phi_1$ and $\alpha$ using standard HAWC software and computed corresponding $TS=2\ln({\cal L}_1/{\cal L}_0)$ values. 
On a second run we computed 95\% confidence level upper limits fixing the spectral index to $\alpha=2.5$ and fitting only $\phi_1$, from where the 95\% confidence level upper limits were estimated using the Feldman-Cousins method~\cite{feldman+cousins}. Upper limits are given here in terms of {\em observed} photon fluxes above 0.5~TeV, which scale directly with the limit on the normalization, $\phi_1^{UL}$, and consider the $\gamma\gamma$ attenuation~(eq.~\ref{horizon}),
$$
N_{obs}(>0.5~{\rm TeV}) \leq \left(4\sqrt{2}\over 3\right) \phi_1^{UL}\cdot{\rm TeV}\, e^{-z/z_h} \, ,
$$
where $z_h=0.108$ as $E_0=0.5~\rm TeV$ and $\alpha=2.5$. Statistical significances are defined as $s=\pm\sqrt{TS}$, where the negative sign refers to sources fitted with $\phi_1\leq 0$. When only one parameter is fitted the distribution of significances $s$ can be approximated by a Gaussian function.  

\subsection{Results}
The first run provided only two detections, the known TeV $\gamma$-ray sources Mrk~421 and Mrk~501. We note that these are also the only two sources detected above $5\sigma$ in the (0.5-2.0)~TeV range with {\em Fermi}-LAT. The search gave a third positive result, a $s=+5.7$ signal on 3FHL~J1652.7+4024, a weak {\em Fermi}-LAT source located in the sky just $0.7^\circ$ from Mrk~501. An inspection of the excess indicates the contamination from Mrk~501. This was one of five sources internally flagged to be considered with particular attention for being within $5^\circ$ of a bright known HAWC source~\cite{2hwc}. 

Markarian~421 was detected with a significance $s=+45.6$ for an intrinsic power law spectrum of index $\alpha=2.42\pm 0.04$ and 1~TeV normalization $\phi_1=(25.7\pm1.2)\times 10^{-12}\,\rm TeV^{-1}cm^{-2}s^{-1}$. 
Markarian~501 was detected with a significance $s=+20.2$ for a spectral index $\alpha=1.96\pm 0.10$ and normalization $\phi_1=(6.8\pm1.1)\times 10^{-12}\,\rm TeV^{-1}cm^{-2}s^{-1}$. The errors quoted are statistical only. A more detailed spectral analysis of these two sources is shown in these proceedings~\cite{mrk-icrc2019}.

\begin{table}
\begin{tabular}{lccccc}
\hline\hline
Object & Class & Redshift & $\phi_1$ & $N_{obs}(>0.5~\rm TeV)$ & Significance \\
 && ($z$)& ($10^{-12} \rm TeV^{-1}cm^{-2}s^{-1}$) & ($10^{-12} \rm cm^{-2}s^{-1}$) & $s$ \\
\hline
Mrk 421 & BL Lac & 0.031 & $28.1\pm 0.7$ & $39.8\pm 1.0$ & +45.8 \\
Mrk 501 & BL Lac & 0.033 & $12.7\pm 0.7$ & $17.6 \pm 0.9$ & +19.9 \\
\hline
M~87 & RDG & 0.004 & $\leq 0.53$ & $\leq 0.95$ & +1.55 \\
IC~310 & RDG$^{\star}$ & 0.019 & $\leq 1.56$ & $\leq 2.47$ & +1.06 \\
1ES~2344+514 & BL Lac & 0.044 & $\leq 5.08 $ & $\leq 6.37$ & +0.70 \\
TXS~0210+515 & BL Lac & 0.049 & $\leq 7.71$ & $\leq 9.24$ & +1.65 \\
1ES~1727+502 & BL Lac & 0.055 & $\leq 3.24 $ & $\leq 3.67$ & --0.40\\
B3~2247+381 & BL Lac & 0.119 & $\leq 2.85 $ & $\leq 1.79$ & --0.64 \\
H~1426+428 & BL Lac & 0.129  & $\leq 8.83$ & $\leq 5.03$ & +0.86 \\
1H~1914--194 & BL Lac & 0.137 & $\leq 24.8$ & $\leq 13.1$ & --0.51 \\
RX~J0648.7+1516 & BL Lac & 0.179 & $\leq 3.56$ & $\leq 1.28$ & --0.51 \\
1H~1013+498 & BL Lac & 0.212 & $\leq 24.8$ & $\leq 6.57$ & --0.11\\
\hline\hline
\end{tabular}
\caption{The twelve best HAWC targets, all fitted assuming $\alpha=2.5$. The upper panel shows the two detections, which are followed by the non-detections, ranked by redshift. The values for the two Markarians assume $\alpha=2.5$, which differs from their best fit value. The limits are 95\% confidence level.}
\label{doce}
\end{table}

\begin{figure}
\includegraphics[width=0.495\textwidth]{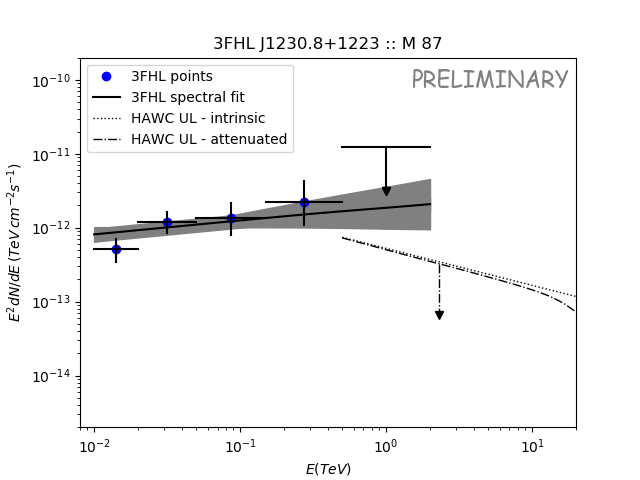}
\includegraphics[width=0.495\textwidth]{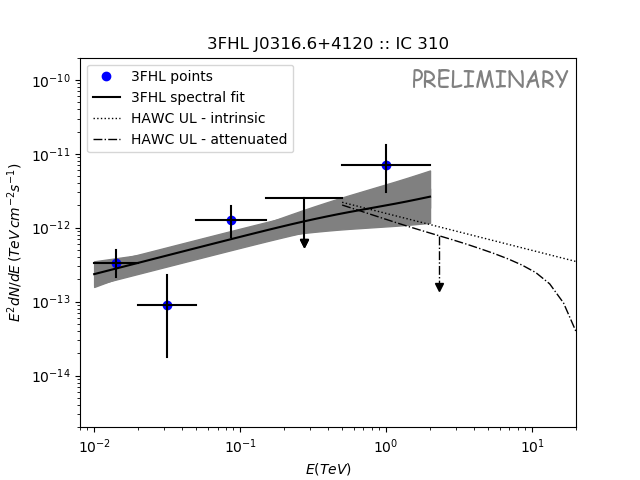}
\caption{High-energy and long term averaged spectra of M~87 ({\em left}) and IC~310 ({\em right}). The points and darker band represent the {\em Fermi}-LAT 3FHL data and spectral fits. The HAWC 95\% CL limits are shown as a dotted line for the intrinsic spectra, with the shaded region including the EBL attenuation.  \label{espectros-rdg}}
\end{figure}

We then computed upper limits in the second run, 
with a fixed spectral index $\alpha=2.5$. In Table~\ref{doce} we show the results for the ten objects identified in section~\S\ref{muestra} which were not detected,
two radiogalaxies and ten BL Lac objects. The radiogalaxies M~87 and IC~310 are of particular interest as their nearer distance make their high-energy spectra mostly unaffected by EBL attenuation. In addition the jet axis of these radiogalaxies does not coincide with the line of sight, providing a different view on the geometry of AGN jet radiation.
In Figure~\ref{espectros-rdg} we show their spectra combining the information in the 3FHL and the HAWC analysis. 

The HAWC limits for M~87 are more than half below the LAT extrapolation in the 0.5--2.0~TeV spectral overlap, and an order of magnitude lower than the 3FHL upper limit in that same band. The HAWC upper limit for M~87 is also below the VHE measurements made in high and intermediate states between 2005 and 2010 with HESS, MAGIC and VERITAS~\cite{m87-vhe}.

IC~310 is a relatively small, but very active, radiogalaxy in the Perseus cluster whose VHE emission has been observed to vary on extremely short timescales, down to {\em minutes}~\cite{ic310-magic}. The LAT spectrum is very hard, with spectral index $\sim 1.5$. This galaxy has a $TS=13$ measurement in the 0.5-2.0~TeV interval by {\em Fermi}, although with non-detection in the 150-500~GeV band. It is observed to be variable in the LAT data. The HAWC upper limit is about a factor of five below the hardest {\em Fermi} data point. Except for that measurement, the HAWC limit may be consistent with a softening of the LAT spectrum. The HAWC upper limit is below the high and flaring states reported by the MAGIC collaboration~\cite{ic310-magic}.

\section{Conclusions and summary}
HAWC detects with high significance the time-averaged TeV emission of Mrk~421 and Mrk~501, the two nearest known BL Lac objects. 
The persistent emission of other active galaxies remains undetected, with photon flux levels one order of magnitude lower than those of the Markarian galaxies, $N(>0.5~{\rm TeV})\lesssim 10^{-12} - 10^{-11}\,\rm cm^{-2}s^{-1}$. Given the limits in sensitivity of air-shower arrays at photon energies below 0.5~TeV, the extragalactic background infrared light is a major obstacle to peer above $z\sim 0.3$. 
In the case of radiogalaxies, like M~87 and IC~310, where $\gamma\gamma$ attenuation is not a major obstacle, the time-averaged emission measured with HAWC is below the extrapolation of the {\em Fermi} spectra reported in the 3FHL catalog, and below high or intermediate activity states measured with ACTs.


\end{document}